\documentclass[]{aa}

\usepackage{graphicx}
\usepackage{txfonts}
\usepackage{mathrsfs}
\usepackage{amsmath,amssymb,amsfonts}
\usepackage{bm}
\newcommand{\msun}{\,\mathrm{M}_\odot}
\newcommand{\gaia}{\emph{Gaia}}
\begin{document}

\title{Masses, bulk densities, and macroporosities of asteroids (15) Eunomia, (29) Amphitrite, (52) Europa, and (445) Edna based on \gaia{} astrometry}

\titlerunning{Masses, densities, and porosities of four large asteroids based on \gaia{} astrometry}

\author{L. Siltala \inst{1} \and M. Granvik\inst{1,2}}

\institute{Department of Physics, P.O. Box 64, FI-00014 University of Helsinki, Finland\\
\email{lauri.siltala@helsinki.fi}
\and
Asteroid Engineering Laboratory, Space Systems, Luleå University of Technology, Box 848, S-98128 Kiruna, Sweden\\
}

\date{Received June 3, 2021; revised September 6, 2021; revised October 25, 2021; accepted October 28, 2021}

 
  \abstract
   {\gaia{} Data Release 2 (DR2) includes milliarcsecond-accuracy astrometry for 14,099 asteroids. One of the main expected scientific applications of these data is asteroid mass estimation via the modeling of perturbations during asteroid-asteroid encounters.}
   {We explore the practical impact of the \gaia{} astrometry of asteroids for the purpose of asteroid mass and orbit estimation by estimating the masses individually for four large asteroids. We use various combinations of \gaia{} astrometry and/or Earth-based astrometry so as to determine the impact of \gaia{} on the estimated masses. By utilizing published information about estimated volumes and meteorite analogs, we also derive estimates for bulk densities and macroporosities.}
   {We apply a Markov chain Monte Carlo (MCMC) algorithm for asteroid mass and orbit estimation by modeling asteroid-asteroid close encounters to four separate large asteroids in an attempt to estimate their masses based on multiple simultaneously studied close encounters with multiple test asteroids. In order to validate our algorithm and data treatment, we apply the MCMC algorithm to pure orbit determination for the main-belt asteroid (367) Amicitia and compare the residuals to previously published ones. In addition, we attempt to estimate a mass for (445) Edna with \gaia{} astrometry alone based on its close encounter with (1764) Cogshall.}
   {In the case of the orbit of (367) Amicitia, we find a solution that improves on the previously published solution. The study of (445) Edna reveals that mass estimation with DR2 astrometry alone is unfeasible and that it must be combined with astrometry from other sources to achieve meaningful results. We find that a combination of DR2 and Earth-based astrometry results in dramatically reduced uncertainties and, by extension, significantly improved results in comparison to those computed based on less accurate Earth-based astrometry alone.}
   {Our mass estimation algorithm works well with a combination of \gaia{} DR2 and Earth-based astrometry and provides very impressive results with significantly reduced uncertainties. We note that the DR2 has a caveat in that many asteroids suitable for mass-estimation purposes are not included in the relatively small sample. This limits the number of asteroids to which mass estimation can be applied. However, this issue will largely be corrected with the forthcoming third \gaia{} data release, which is expected to allow for a wave of numerous accurate mass estimates for a wide range of asteroids.}


   \maketitle
%

\section{Introduction}

We present and discuss the masses and densities of four large asteroids -- (15) Eunomia, (29) Amphitrite, (52) Europa, and (445) Edna. For the first time ever, \gaia{} astrometry of asteroids \citep{DR2,DR2_Fed} is used for estimating asteroid masses, a long-anticipated contribution made possible by the \gaia{} mission \citep{Mig07,Mou08,DR1}.

An asteroid's mass can be seen as a fundamental property that, when combined with a volume estimate, allows for a trivial computation of the asteroid's bulk density. Combining the bulk density with, on one hand, compositional information of the asteroid obtained primarily through spectroscopic observations and, on the other hand, spectra and densities of meteorites found on the Earth allows the asteroid's bulk composition and overall structure to be constrained \citep{Bri02,Car12}. Hence, it is clear that knowledge of an asteroid's mass alongside the volume is critical for essentially all detailed studies of the interior characteristics of asteroids.


Asteroid mass estimation is traditionally performed via the measurement and modeling of orbital perturbations caused by a massive asteroid on a smaller test asteroid, which is usually assumed to be massless, during a close encounter between the two. In practice, this presents an at least 13-dimensional inverse problem where the aim is to simultaneously fit a six-dimensional Cartesian state vector (or another set of parameters that describe the orbit) for each asteroid in addition to the mass of the perturber by accurately reproducing the astrometry for both objects over a long time span, and including observations both before and after the close encounter. Different close encounters with multiple test asteroids may be used simultaneously, which significantly reduces the uncertainty on the resulting mass \citep[see, e.g.,][]{Bae17,Sil20} at the cost of additional computational complexity. 

We note that alternative approaches for mass estimation also exist. It is possible to measure asteroidal perturbations upon the orbits of the planets, Mars and the Earth in particular. Most recently, this approach was used by \citet{Fie20} to estimate the masses of 103 asteroids with uncertainties of less than 33\%. Alternatively, in the case of binary or triple asteroids, the components of which have well-constrained orbits, it is possible to utilize Kepler's third law to estimate the mass of the system. This approach was most recently used to estimate a new mass for (216) Kleopatra \citep{Bro21}. Equivalently, an asteroid's gravity field and thus mass may be measured by spacecraft flyby or orbiter data, which provide the most accurate mass estimates by far. This approach was most recently applied to (162173) Ryugu based on Hayabusa2 measurements \citep{Wat19} and to (101995) Bennu based on OSIRIS-REx measurements \citep{Lau19}.

The wealth of data included in \gaia{} Data Release 2 (DR2) includes astrometry of unprecedented milliarcsecond-scale accuracy for 14,099 asteroids \citep{DR2}. Asteroid mass estimation via the modeling of asteroidal perturbations during asteroid-asteroid close encounters is noted as one of several practical applications for the Solar System object observations in the DR2 \citep{DR2_Fed}. The challenges in the field arise from the fact that, for all but the largest asteroids, the masses in question are small compared to planetary masses, and, by extension, the perturbations are weak and often difficult to measure with decent accuracy. The practical advantages of high-accuracy \gaia{} astrometry are thus clear: more accurate astrometry will lead to more accurate results and thus constrain the uncertainties of the mass and the orbits in comparison to estimates computed with Earth-based data alone. The accurate \gaia{} astrometry will likely also allow the estimation of the masses of certain asteroids that still lack mass estimates due to the perturbation signal being too weak to stand out in the residuals between the observed and theoretically predicted positions of less accurate preexisting observations.

In this study we demonstrate, for the first time, the practical impact of DR2 astrometry on asteroid mass estimation. We first verify that our algorithm and data treatment works as expected by applying it to the task of computing an orbit for (367) Amicitia and comparing the resulting residuals to those reported by \citet{DR2_Fed}. We then test the combined orbit and mass estimation, and present and compare three separate results for the mass of asteroid (445) Edna based on a close encounter with asteroid (1764) Cogshall with DR2 data alone, with Earth-based data alone, and with a combination of the two. Finally, we apply our methods to several large asteroids by combining DR2 data with Earth-based data and compare our results to equivalent results computed with Earth-based data alone.

\section{Observational data and data treatment}
\label{data_treatment}
We used all astrometry available through the Minor Planet Center as of 14 January 2021 for our target asteroids. We corrected for systematic offsets in the astrometry induced by biases in old astrometric star catalogs with a recent model based on the DR2 star catalog (P. Tanga, private communication; \citealt{Tan20}).
We rejected astrometry that could not be debiased due to, for example, a lack of information on which star catalog was used in the astrometric reduction. Notably, this led to the rejection of a large amount of older data, pre-1990 data in particular. The data were weighted using the statistical per-observatory uncertainties as provided by a recent error model (F. Spoto, private communication; \citealt{Fer20b,Tan20}).
The uncertainties were multiplied by a factor of $\sqrt{N}$ for $N$ same-night observations of a single target from a single observatory to account for otherwise un-modeled correlations between the same-night observations \citep{Far15}. The chosen approach is known to improve the results obtained with the aforementioned error model (F.~Spoto, private communication).
The formal weights of an individual Earth-based observation, as later shown in Eq.~\ref{chi2_definition}, are the inverses of the corresponding variances, namely $1/(N\sigma^2),$ where $\sigma$ is the observation's astrometric uncertainty provided by the error model.
Henceforth, we refer to this data set as the Earth-based data.

The DR2 data were largely used as is, with right ascension (RA) and declination (Dec) coordinates directly obtained from the official data release. We used each observation available for our target asteroids included in the DR2. We constructed covariance matrices for the observations from the standard deviations and correlations provided in the DR2 to serve as weights for each observation. A known caveat with the DR2 is that a relativistic light-bending correction was erroneously applied to the astrometry of objects in the Solar System as if they were infinitely distant objects (i.e., stars). We reversed this erroneous correction with code graciously provided by F.~Mignard. In our experience, the magnitude of this effect is tiny even in comparison to the accuracy of \gaia{} astrometry and thus does not have a significant impact on the results. We applied the correction nonetheless as it remains the known DR2 issue that is expected to be corrected in future data releases. 
Table~\ref{data} shows the number of Earth-based observations available for each asteroid both pre- and post-debiasing in addition to the number of DR2 observations available. Statistics as to the quality of the data are also shown in terms of the non-weighted root mean square (rms) values for the residuals in RA and Dec for Earth-based data, and in terms of along-scan (AL) and across-scan (AC) for \gaia{} astrometry corresponding to maximum-likelihood solutions, in addition to the number of detected outliers for both. \gaia{} data can be seen to reach their expected accuracy for all objects, but we note that for (124)~Alkeste our model finds an unusually high number of outliers in \gaia{} data.

\begin{table*}
  \begin{center}
  \caption{Statistics for both the Earth-based and the \gaia{} astrometry of each object.}\label{data}
    \begin{tabular}{cccccccccc} 
      \hline 
      \hline 
      Asteroid    & RMS (RA)   & RMS (Dec) & $N_{tot}$ & $N_{use}$ & $N_{out}$ & RMS (AL) & RMS (AC) & $N_{DR2}$ & $N_{DR2out}$ \\
                  & [as]       & [as]      &            &           &           & [mas]    & [as]     &           &              \\ 
      \hline
      \textbf{367}    & 0.42 & 0.36 & 2888 & 2260 & 18 & 0.72 & 0.25 & 132  & 4   \\
      15              & 0.66 & 0.50 & 2800 & 1198 & 67 & N/A  & N/A  & N/A  & N/A \\
      \textbf{1537}   & 0.52 & 0.48 & 2914 & 1989 & 19 & 0.74 & 0.15 & 147  & 2   \\
      \textbf{13724}  & 0.48 & 0.49 & 2133 & 1594 & 4  & 2.07 & 0.54 & 296  & 9   \\
      \textbf{2671}   & 0.43 & 0.39 & 2903 & 2596 & 27 & 0.98 & 0.15 & 88   & 1   \\
      50278           & 0.51 & 0.50 & 812  & 689  & 1  & N/A  & N/A  & N/A  & N/A \\
      411232          & 0.68 & 0.55 & 162  & 80   & 0  & N/A  & N/A  & N/A  & N/A \\
      29              & 0.62 & 0.57 & 2612 & 1107 & 34 & N/A  & N/A  & N/A  & N/A \\
      \textbf{362}    & 0.52 & 0.45 & 3115 & 2128 & 20 & 0.70 & 0.45 & 340  & 36  \\
      \textbf{987}    & 0.46 & 0.35 & 2914 & 2088 & 24 & 0.83 & 0.27 & 180  & 17  \\
      \textbf{9741}   & 0.53 & 0.43 & 2934 & 1566 & 13 & 1.61 & 0.47 & 103  & 1   \\
      43142           & 0.48 & 0.45 & 1759 & 1194 & 1  & N/A  & N/A  & N/A  & N/A \\
      77424           & 0.55 & 0.49 & 703  & 520  & 6  & N/A  & N/A  & N/A  & N/A \\
      52              & 0.51 & 0.50 & 3436 & 2010 & 34 & N/A  & N/A  & N/A  & N/A \\
      \textbf{124}    & 0.52 & 0.51 & 3621 & 2345 & 36 & 0.7  & 0.22 & 84   & 51  \\
      \textbf{627}    & 0.55 & 0.42 & 3338 & 2330 & 20 & 0.64 & 0.15 & 95   & 0   \\
      \textbf{8269}   & 0.50 & 0.44 & 2074 & 1129 & 3  & 2.3  & 0.24 & 184  & 0   \\
      81049           & 0.48 & 0.53 & 1061 & 784  & 3  & N/A  & N/A  & N/A  & N/A \\
      14723           & 0.48 & 0.48 & 1726 & 1142 & 1  & N/A  & N/A  & N/A  & N/A \\
      \textbf{445}    & 0.44 & 0.38 & 2711 & 1882 & 28 & 0.84 & 0.20 & 300  & 44  \\
      \textbf{1764}   & 0.48 & 0.45 & 3296 & 2395 & 14 & 1.00 & 0.18 & 255  & 22  \\
      \textbf{5104}   & 0.46 & 0.40 & 2681 & 1950 & 14 & 1.08 & 0.33 & 121  & 3   \\
      71031           & 0.45 & 0.44 & 1048 & 659  & 0  & N/A  & N/A  & N/A  & N/A \\
      \hline 
    \end{tabular}
    \tablefoot{The columns represent the rms values
    in terms of RA and Dec for the Earth-based astrometry, the total number of Earth-based astrometric data points obtained from the Minor Planet Center for each asteroid, the number of debiased astrometric data points used in the computation, the number of rejected outliers in the Earth-based data, and the same statistics in terms of AL and AC for the \gaia{} DR2 astrometry. The values are computed based on residuals corresponding to maximum-likelihood MCMC proposals. Asteroids whose numbers are written in boldface are included in the DR2.}
  \end{center}
\end{table*}

A challenge in using the DR2 for asteroid mass estimation is the relatively low number of asteroids for which astrometry was included (14,099). At first glance it may seem a large number, but most massive asteroids are excluded from the DR2, as are many interesting test asteroids, which limits the number of cases the DR2 can be applied to. For this work, we opted to estimate the masses of four asteroids: (15) Eunomia, (29) Amphitrite, (52) Europa, and (445) Edna. For all but (445) Edna we performed four separate mass estimations with two sets of test asteroids, whereas for Edna seven mass estimations were performed with three different sets. For each set of test asteroids, separate runs were performed with and without all available DR2 data in order to appropriately gauge the impact of the DR2 data on the results. The chosen perturbers and test asteroids are listed in Table~\ref{results}.



Asteroid (445) Edna is a special case as it had a known close encounter with (1764) Cogshall on 31 October 2014 \citep{Gal02}, which roughly corresponds to the middle of the DR2 observational time span. Additionally, both Edna and Cogshall are included in the DR2. Thus, as DR2 data are available both before and after the close encounter, it is possible to attempt mass estimation for (445) Edna based on the DR2 data alone. We estimated Edna's mass with Cogshall as the sole test asteroid in three separate scenarios: with the DR2 data alone, with the Earth-based data alone, and with a combination of both.

\section{Methods}

\subsection{Robust adaptive Metropolis algorithm for mass estimation}

The nonlinear inverse problem of determining an asteroid's mass has typically been solved with linearized least-squares methods. The methods used have the inherent limitation that one needs to make assumptions regarding the shape of the underlying probability distribution, namely Gaussian uncertainties for the resulting masses and orbits are assumed. However, \citet{Sil17} demonstrated that the uncertainties are not necessarily Gaussian in cases with large uncertainties, which casts a certain amount of doubt on such assumptions. We have also observed that our algorithm provides uncertainties closer to Gaussian when more and/or higher precision data are included in the model, for example in the form of additional test asteroids \citep{Sil20}.
It is also widely suspected that uncertainties in asteroid masses tend to be significantly underestimated, which is suggested by the fact that different mass estimates for a single asteroid reported in the literature, sometimes even from the same study, strongly contradict one another \citep{Bri02,Car12}.

In this work, we applied our Markov chain Monte Carlo (MCMC) mass estimation algorithm, which is primarily based on the robust adaptive Metropolis (RAM) scheme \citep{Vih12}. The algorithm is documented in greater detail in \citet{Sil20}, and a version with minor improvements was recently applied to asteroid (16)~Psyche \citep{Sil21}. 

For the present work we made further improvements to the numerical method. A stricter outlier rejection criterion was used in which observations with a Mahalanobis distance $d_\mathrm{M} \ge 3$ are rejected, whereas previously we had rejected those with $d_\mathrm{M} \ge 4$. The change was made after we observed that poorly fitting observations with large $d_\mathrm{M}$ were adversely impacting both the results and the numerical stability of the MCMC chain when the DR2 astrometry was included. Other aspects of the outlier rejection procedure remain the same as in \citet{Sil20,Sil21}. For example, (i) both coordinates of observations rejected as outliers are excluded from the analysis, and (ii) a previously rejected observation may later be re-included if the outlier-rejection procedure no longer classifies the observation in question as an outlier.

We also significantly reduced the width of the initial proposal distribution for the mass as it was much too wide to be realistic when including the DR2 astrometry. The change optimizes the initial burn-in phase but does not noticeably impact the final results as the RAM scheme will nonetheless eventually adapt the proposal distribution to optimal values.

All of the computations were performed with the publicly available open-source asteroid-orbit-computation software OpenOrb\footnote{https://github.com/oorb/oorb} \citep{Gra09} with the corresponding code included in the 1.2.0 release. The perturbations of Pluto, the Moon, and the eight planets were included in the force model via the DE430 planetary ephemerides \citep{Fol14}, whereas the perturbations of the five most massive asteroids, (1)~Ceres, (2)~Pallas, (4)~Vesta, (10)~Hygiea, and~(704) Interamnia, were included via the BC430 asteroid ephemerides \citep{Bae17}. Together, these five asteroids contain more than half of the total mass of the asteroid belt. The force model could be further improved by including a larger number of BC430 asteroids, but we settled for the largest five due to computational constraints and expected diminishing returns from including asteroids with lower masses. Such asteroids also tend to have larger mass uncertainties, which are unaccounted for in BC430. We also verified that the target asteroids do not have significant gravitational interactions with other BC430 asteroids through the application of the close encounter finding algorithm described in the following section.

\subsection{Algorithm to find close encounters}

During our previous mass estimation studies, we relied solely on close encounters that had been previously documented and, in many cases, studied in practice. Relying on such studies alone comes with certain drawbacks, however: it is often not clear which close encounters are optimal for use in the sense that the signal of the perturbations as seen in the astrometric residuals is particularly strong. Past studies focused on finding close encounters suitable for asteroid mass estimation, such as \citet{Gal02} and \citet{Fie03}, often report the dates of closest approach and minimum distances between the asteroids in addition to the relative encounter velocity. That information allows one to make a preliminary assessment of the usefulness of different encounters. However, there is no consideration of the observational data available for the asteroids in question. It is conceivable that a close encounter that produces a weaker signal can in practice be superior to an encounter that provides a stronger signal if the former case has superior astrometry in terms of quality and/or quantity available.

On the other hand, certain prior mass estimation studies \citep[e.g.,][]{Bae17} report asteroid masses and their uncertainties individually for each close encounter studied in addition to results based on combinations of those encounters. One can then examine such preexisting results and the corresponding uncertainties of each encounter. It seems reasonable to assume that the close encounters that previous studies found to result in the lowest uncertainty of the mass should also provide excellent results in other studies. However, these results naturally cannot include observations obtained after the publication of the study in question, and thus the results are, to some extent, bound to become outdated over time.

Finally, we have observed that different studies have a tendency to find entirely different close encounters, which led us to suspect that the methods used may not produce complete lists of close encounters. As a demonstrative example, \citet{Fie03} reported as many as 29 close encounters involving (16)~Psyche, while \citet{Gal02} reported nine; furthermore, there was not a single close encounter simultaneously detected by both studies. At the same time, \citet{Bae17} more recently used eight separate close encounters to study the mass of (16)~Psyche, none of which had previously been reported by \citet{Fie03} and only one of which had been reported by \citet{Gal02}. In addition, for most asteroids very few or no encounters whatsoever have been reported in the literature.

In an attempt to remedy these issues, we present a new three-stage approach built on OpenOrb to search for close encounters. The approach is to first integrate the orbits of all known asteroids both into the past and into the future across a desired time span, which results in a sequence of Cartesian state vectors as a function of time for each asteroid. We applied OpenOrb's $n$-body integrator to this purpose with a force model equivalent to that used in the mass estimation process as described in the previous section.
Although it is a computationally intensive step, it does not have to be repeated until one wants to update the initial orbits. 
Next, a short Python script is utilized to analyze the state vectors by computing Euclidean distances as a function of time between a given perturbing asteroid and every other asteroid in the data set to find the asteroids with the smallest minimum distances to the perturbing asteroid. 

In the third stage, the script downloads all currently available astrometry for the asteroids in question from the Minor Planet Center and uses these data to apply the marching mass-estimation algorithm found in OpenOrb \citep{Sil17}. The marching algorithm reduces the problem to one dimension by sampling across a range of possible perturber masses and recording the goodness of fit for each mass in terms of the $\chi^2$ of the test asteroid. The approximation is hardly sufficient to obtain scientifically useful mass estimates, but it does show the strength of the signal of the perturber's mass with only a modest computational cost. The larger the $\chi^2$ difference across the range of the masses (i.e., $\Delta\chi^2 = \chi^2_{\mathrm{max}} - \chi^2_\mathrm{min}$, where $\chi^2_{\mathrm{max}}$ and $\chi^2_\mathrm{min}$ refer to the maximum and minimum values), the stronger the signal of the perturbation is, based on the available data. Thus, through analyzing the correlation between the perturber mass and the goodness of fit $\chi^2$, the script detects and returns a list of close encounters for the desired perturbing asteroid, sorted by the strength of the signal in each encounter, and this list can be then directly used to select an optimized set of close encounters.  For an individual asteroid, $j$, with $N_\mathrm{obs}$ observations, $\chi^2$ is defined in matrix notation as
\begin{equation}
\label{chi2_definition}
\chi^2 = \sum_{i=1}^{N_{\mathrm{obs}}} 
\left[\bm{\epsilon}^T_{i,j}\bm{\Sigma}^{-1}_{i,j}\bm{\epsilon}_{i,j}\right]
.\end{equation}
Here, $\bm{\Sigma}_{i,j}^{-1}$ is the the inverse of a given observation's covariance matrix, also called the information matrix, of a given observation, and $\bm{\epsilon_{i,j}}$ is a vector that consists of the corresponding observation's observed minus computed ($O-C$) residuals.

Based on tests, the algorithm works remarkably well and indeed finds many previously reported close encounters in addition to many more that have not been previously reported to the best of our knowledge. The results are used in this work both to select the most useful test asteroids from those previously reported and, where necessary or useful, to find new encounters suitable for mass estimation. A more detailed analysis and a thorough report of useful close encounters across a larger number of perturbing asteroids remains to be done. We note that the OpenOrb part of the code is included in the official OpenOrb repository, but the Python script is not. The Python script is, however, available from the corresponding author upon request.

\section{Results and discussion}

\subsection{Orbit of (367) Amicitia}

To verify that both the algorithm and the data treatment function correctly with the DR2 data, we only computed the orbit of main-belt asteroid (367) Amicitia with the MCMC algorithm. That is, its mass was not considered at all. Amicitia was used earlier as an arbitrarily chosen example that highlighted the accuracy of the DR2 \citep{DR2_Fed}. For this particular asteroid, preexisting published results thus are available from the \gaia{} Collaboration to compare ours against.

\begin{figure} 
  \begin{center} 
    \includegraphics[width=1.0\columnwidth]{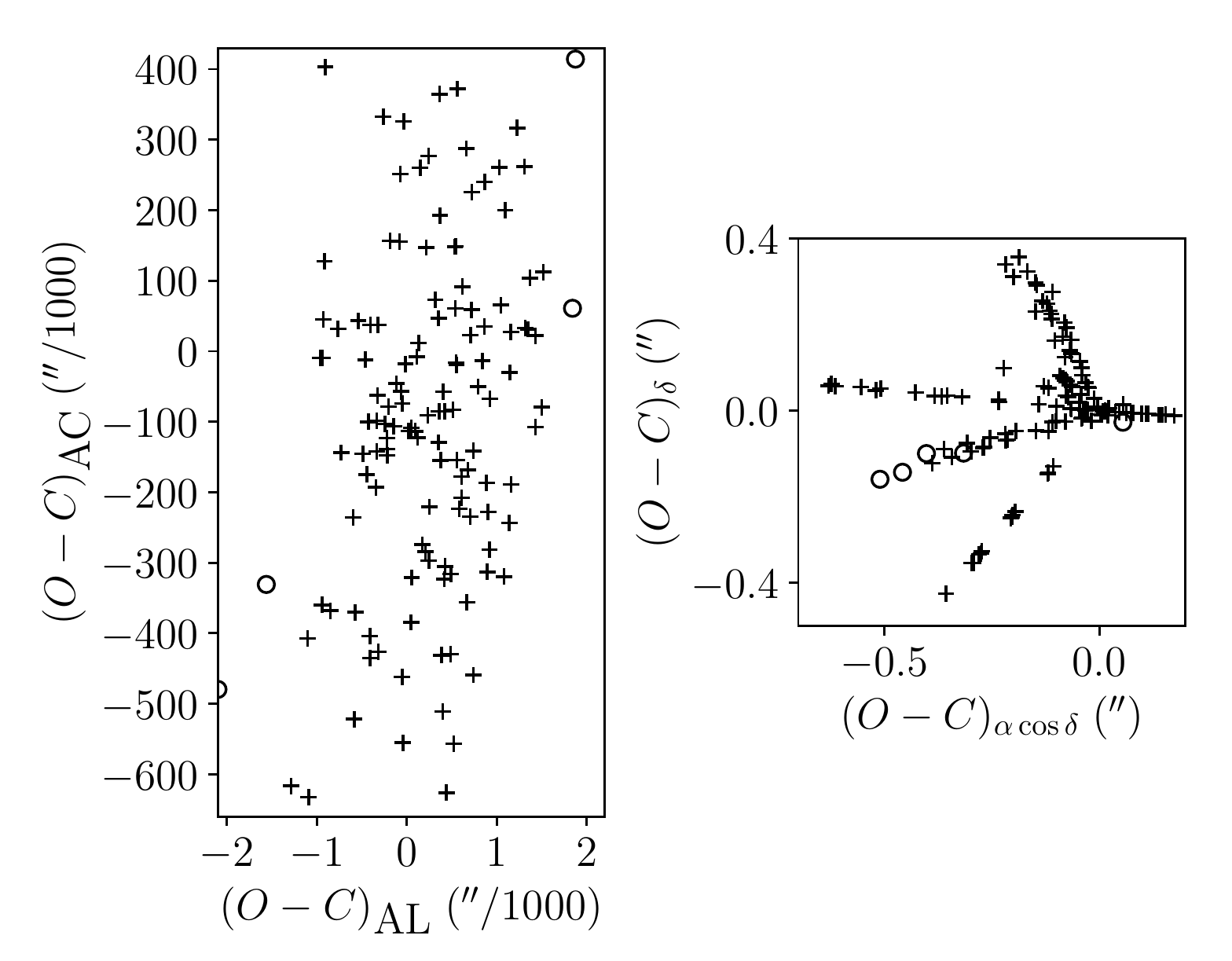}
    \caption{Residuals of the DR2 observations for (367) Amicitia corresponding to the maximum-likelihood MCMC solution, computed using the DR2 astrometry alone, in terms of AL and AC coordinates (left), and in terms of RA ($\alpha$) and Dec ($\delta$) (right). The unfilled circles represent data rejected as outliers. The scales of the plots have intentionally been chosen to match Fig.~23 of \citet{DR2_Fed} and thus allow for direct visual comparison.}
    \label{amicitia_resids}
  \end{center}
\end{figure}

The observed minus computed $(O-C)$ residuals corresponding to the maximum-likelihood solution are shown in Fig.~\ref{amicitia_resids} in terms of RA and Dec as well as AL and  AC coordinates on the \gaia{} focal plane. Upon visual inspection and comparison to Fig.~22 in \citet{DR2_Fed}, it is clear that the RA and Dec residuals are essentially the same, as expected. However, upon applying the transformation to AL and AC coordinates \citep[Eq.~3 in][]{DR2_Fed} and comparing the results to Fig.~23 in \citet{DR2_Fed}, a clear difference can be seen, and the distributions of the residuals no longer match. The implication is that the resulting orbits are not identical. The discrepancy is not necessarily of concern; an analysis of our AL residuals yields $1\sigma$ limits of 0.71~mas, which is well within expectations for DR2 data. \citet{DR2_Fed} unfortunately does not report precise numbers for their $1\sigma$ limits obtained for Amicitia's residuals, but visual inspection of the $1\sigma$ boundaries in their Fig.~23 gives an equivalent value of approximately 0.86~mas with the exact same data set. This can also be seen visually upon comparison of the figures. For instance, Fig.~23 in \citet{DR2_Fed} shows 15 AL residuals smaller than -1~mas, whereas our Fig.~\ref{amicitia_resids} shows only five such AL residuals. Thus, at least for this test case, the MCMC algorithm finds an orbit solution that more accurately reproduces the observed positions. In terms of AC, an apparent negative systematic bias can be seen in the residuals. This is a known feature of DR2 astrometry of asteroids and was also observed in the DR2 performance verification paper \citep{DR2_Fed}, according to which it is explained by the residuals following the direction of \gaia{}'s velocity vectors. In both cases, the resulting orbits are shown in Table~\ref{amicitia_table}. We used the observational mid-epoch as the default inversion epoch for the orbit. The results of Amicitia with the simultaneous use of the \gaia{} and Earth-based astrometry were the only exception to the default behavior so as to allow for a direct comparison of two orbits that are based on two different data sets.

\begin{figure}  
  \begin{center} 
    \includegraphics[width=1.0\columnwidth]{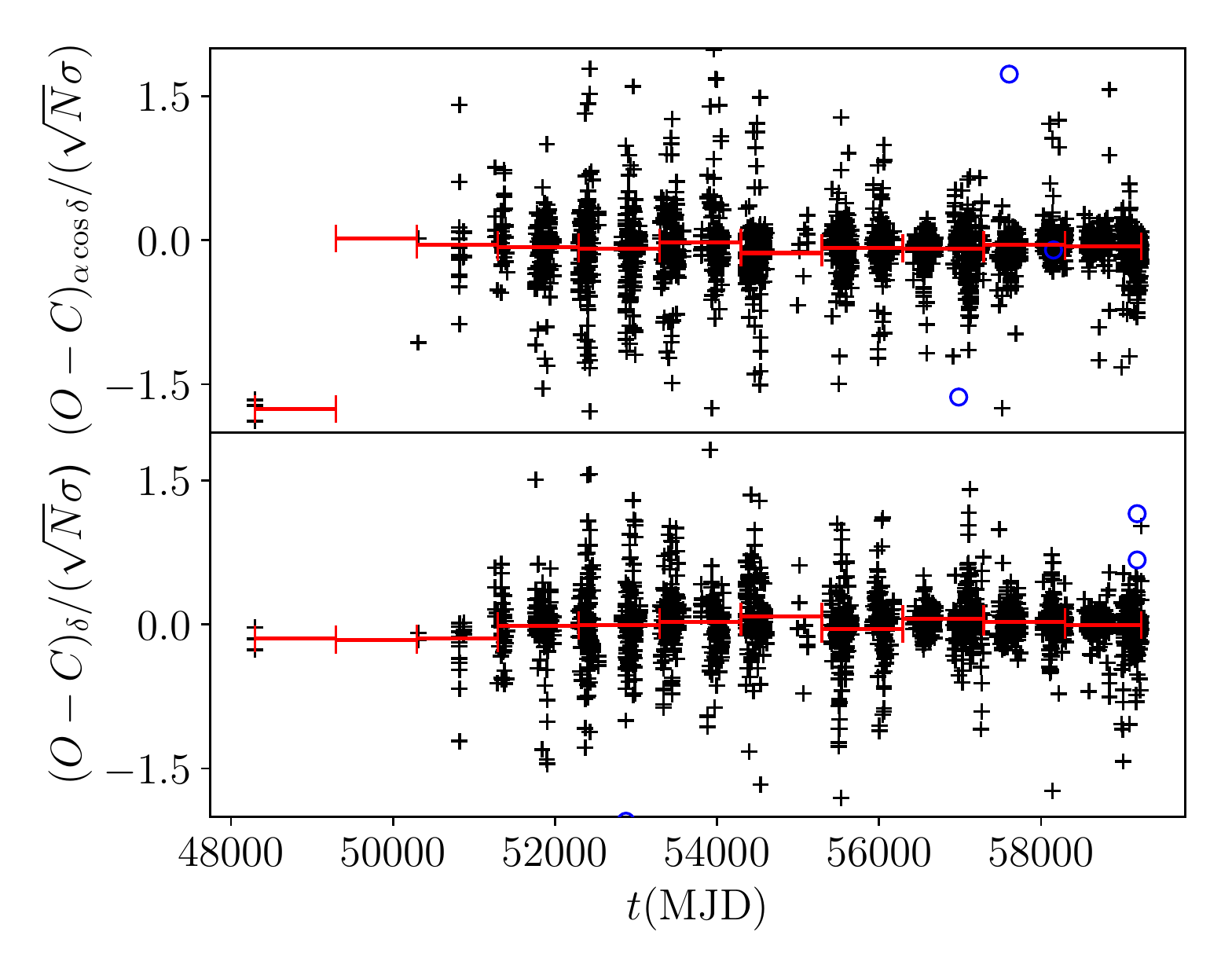}
    \caption{Residuals of the Earth-based observations for (367) Amicitia corresponding to the maximum-likelihood MCMC solution computed using a combination of the DR2 and the Earth-based astrometry, as a function of modified Julian date. The residuals of each observation are divided by their assumed uncertainties,$\sqrt{N}\sigma$, as described in Sect.~\ref{data_treatment}. The black data points represent the Earth-based astrometry, and the blue circles represent astrometry rejected as outliers. The red horizontal lines represent rolling means of the residuals computed over consecutive 1000-day periods. The error of the mean is indistinguishable on this scale. Several data points are not visible due to the y-axis scale chosen for the figure. Consequently, some outlier observations in particular may only be visible in one of the two coordinates.} \label{amicitia_gaiaandearth}
  \end{center}
\end{figure}

Next, we recomputed Amicitia's orbit with a combination of the DR2 and the Earth-based astrometry to gauge the performance of our algorithm. 
Figure~\ref{amicitia_gaiaandearth} shows the residuals in RA and Dec, divided by their assumed uncertainties, over time for the entire Earth-based data set that corresponds to the maximum-likelihood solution. Analysis of the distribution of the residuals yields a median value of -0.06 arcseconds for RA and 0.01 arcseconds in Dec, which are reasonably good results; however, they suggest that there is a small systematic negative bias, especially in RA.
A simple signs test (i.e., $M = (N(+) - N(-))/2$, where $N(+)$ is the number of positive and $N(-)$ the number of negative residuals) yields $M_{\alpha\cos\delta} = -466$ and $M_\delta = 136$. With the null hypothesis of $M$ being zero, that is, the residuals being evenly distributed between positive and negative values, the results correspond to $p$ values of  $8 \times 10^{-69}$ in RA and $4 \times 10^{-7}$ in Dec, which means that the signs are not evenly distributed.
To further examine the matter, we computed rolling means for the residuals divided into consecutive 1000-day slices, which are overlaid on Fig.~\ref{amicitia_gaiaandearth} as red horizontal lines. These means are mostly negative, as expected based on the sign test, but are smaller than one-tenth of the assumed random uncertainties. Thus, the systematic effect is very small relative to the assumed random uncertainties of the observations. The first three observations have a significant offset in RA. They were taken by the same observatory on the same night and thus have weights reduced by $\sqrt{3}$. This points to a problem with the timing of the observations. Nonetheless, the errors were small enough that the observations were not rejected by our outlier detection algorithm.
Furthermore, based on our testing, the small residual trend largely disappears when only ground-based data are used. Thus, the trend may be a consequence of the known, previously discussed systematic bias in the AC direction in \gaia{} DR2.
As the residual trends are marginal compared to the random uncertainties, and thus unlikely to be a significant issue, we leave the matter to be investigated further in future work with \gaia{} Data Release 3 (DR3), which will include an order of magnitude more data and thus allow for a more detailed study of the systematics.

\begin{figure}   
  \begin{center} 
    \includegraphics[width=1.0\columnwidth]{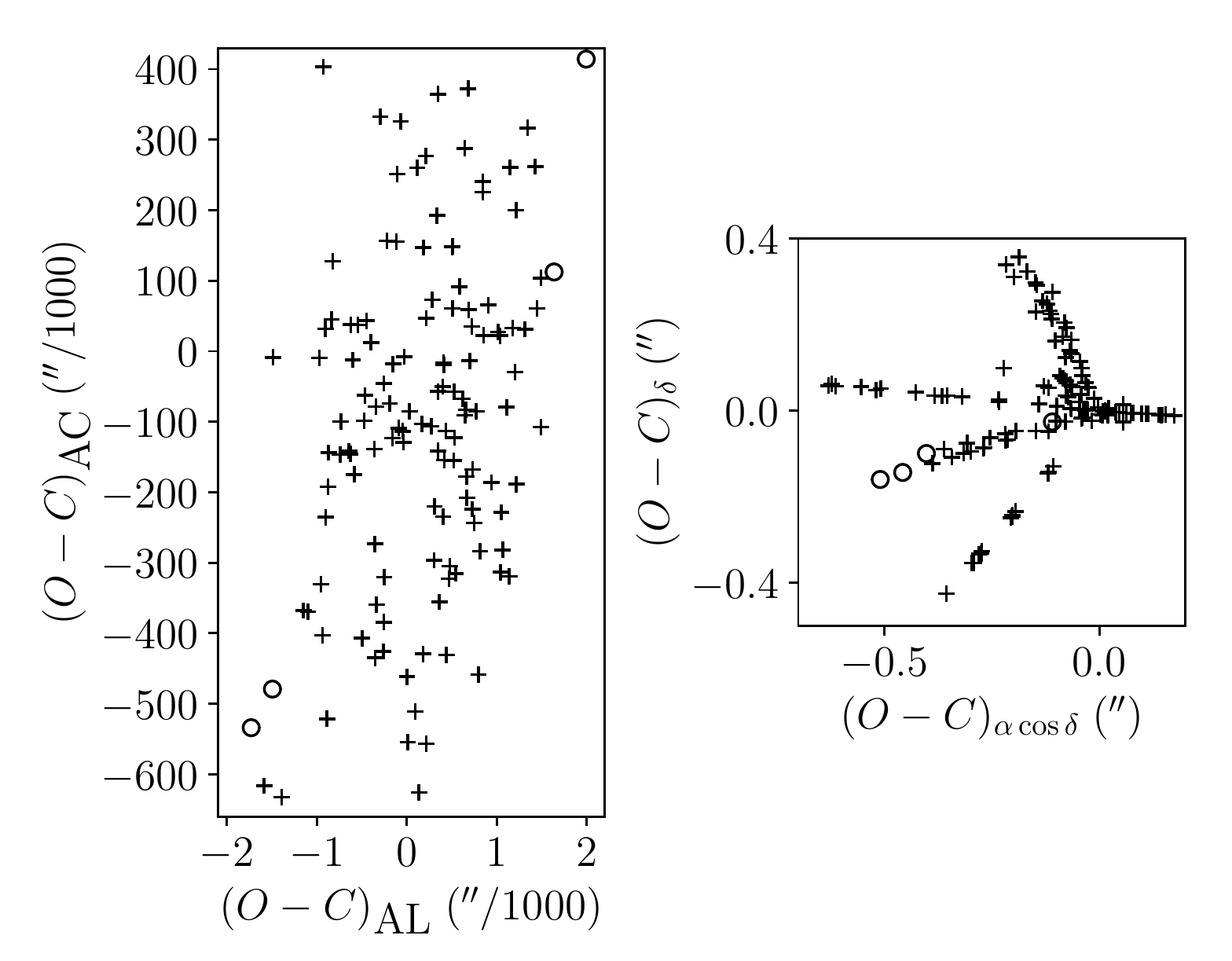}
    \caption{As in Fig.~\ref{amicitia_resids} but based on a MCMC run that includes both the DR2 and the Earth-based astrometry.}
    \label{amicitia_gaia}
  \end{center}
\end{figure}

Residuals of the \gaia{} observations are shown separately in Fig. ~\ref{amicitia_gaia}, which again shows that the milliarcsecond accuracy \gaia{} promises is more clearly visible in the (AL, AC) plane. In this case, the $1\sigma$ boundaries of the residuals in AL are 0.76~mas, in comparison to 0.71~mas for the fit with the DR2 astrometry alone. Thus, the residuals corresponding to the DR2 astrometry have worsened somewhat in comparison to orbits computed with the DR2 astrometry alone, which can be explained by the orbit having to simultaneously fit both the DR2 and the Earth-based astrometry. Even so, the fit remains better than the DR2-only fit reported in \citet{DR2_Fed}, which has corresponding $1\sigma$ boundaries of 0.86~mas.

We conclude that our orbital solution for (367) Amicitia reveals no issues with our algorithm but rather suggests that it is an improvement compared to the solution presented by \citet{DR2_Fed} in terms of $(O-C)$ residuals of \gaia{} observations.

\begin{table*}
  \begin{center}
    \caption{Osculating orbital elements for (367) Amicitia.}\label{amicitia_table}
    \begin{tabular}{lcccccc} 
      \hline
      \hline
      & $a$ (au) & $e$ & $i$ (deg) & $\Omega$ (deg) & $\omega$ (deg) & $M$ (deg) \\
      \hline
       DR2 and the Earth  & $\phantom{\pm}2.2191846387$ & $\phantom{\pm}0.09570674706$ & $\phantom{\pm}2.941623063$ & $\phantom{\pm}83.45773139$ & $\phantom{\pm}55.48997013$ & $\phantom{\pm}59.82257272$ \\
                      & $\pm 0.0000000019$ & $\pm 0.00000000086$ & $\pm 0.000000015$ & $\pm \phantom{0}0.00000050$ & $\pm \phantom{0}0.00000072$ & $\pm \phantom{0}0.00000051$ \\
       DR2 & $\phantom{\pm}2.2191846545$ & $\phantom{\pm}0.09570674562$ & $\phantom{\pm}2.941623059$ & $\phantom{\pm}83.45773182$ & $\phantom{\pm}55.4899737\phantom{0}$ &  $\phantom{\pm}59.8225690\phantom{0}$ \\
           & $\pm 0.0000000074$ & $\pm 0.00000000088$ & $\pm 0.000000015$ & $\pm \phantom{0}0.00000053$ & $\pm \phantom{0}0.0000019\phantom{0}$ &$\pm \phantom{0}0.0000018\phantom{0}$ \\
      \hline
    \end{tabular}
    \tablefoot{The elements are given in the heliocentric frame and correspond to the epoch MJD 58174.0 TT, which is the observational mid-epoch of the DR2 observations used rounded to the nearest midnight.}
  \end{center}
\end{table*}

\subsection{Asteroid volumes utilized for bulk density computation}
Discussion on the volume-equivalent diameters we utilize is warranted here, as computation of bulk densities obviously requires knowledge of the volume in addition to that of the mass.  We have observed that, for our target asteroids, some discrepancy can be seen between the volume-equivalent diameters reported by separate individual studies; as an example, \citet{Rya10} report two separate diameters of $81.40 \pm 4.34$ and $98.178 \pm 5.30$ kilometers for (445)~Edna, one of the asteroids studied in this paper, whereas \citet{Mas11} and \citet{Usu11} respectively report diameters of $89.16 \pm 1.42$ and $105.50 \pm 1.51$ kilometers for the same asteroid. Evidently, there are very strong disagreements between the different studies, which suggests underestimated uncertainty. In fact, past studies also raise similar concerns \citep{Han17,Usu14} regarding the reliability of the uncertainties of asteroid diameters in the literature in a broader sense.

Considering these issues, we opted to use the diameters provided by the SiMDA (Size, Mass and Density of Asteroids) database \citep{Kre20} throughout, which are averages computed using the expected value method based on values obtained from multiple separate studies. This method provides larger uncertainties than, for example, a weighted average, but it is in fact believed that these larger uncertainties are more realistic and robust to outliers \citep{Bir14}. In comparison to the previously cited values, \citet{Kre20} reports a diameter of $86.3 \pm 8.28$ kilometers for (445)~Edna; here, the uncertainties are, evidently, significantly wider. In conclusion, with a different selection of asteroid diameters we would correspondingly obtain lower uncertainties for bulk densities, but the reliability of such estimates would be questionable to a certain extent. As such, we chose to use conservative estimates for asteroid diameters while noting for interested readers that it is trivial to recompute the corresponding densities with different values for the diameter.

\subsection{Mass and density of (445) Edna}
\begin{table*}
  \begin{center}
  \caption{Compilation of the MCMC mass estimation results.}\label{results}
    \begin{tabular}{cccccc} 
      \hline
      \hline
      Perturber    & Test asteroids   & Mass w/ Gaia       & Mass w/o Gaia      & Diameter & Density           \\
                   &                          & [$10^{-11} \msun$] & [$10^{-11} \msun$] & km       & g/cm$^3$  \\ 
      \hline
      15           & \textbf{1537},\textbf{2671},\textbf{13724}                 & $1.32^{+0.21}_{-0.16}$       & $0.97^{+0.51}_{-0.49}$       & $255 \pm 13$ & $3.01^{+0.59}_{-0.66}$ \\[0.1cm]
      15           & \textbf{1537},\textbf{2671},\textbf{13724},50278,411232    & $1.522^{+0.047}_{-0.051}$     & $1.36^{+0.45}_{-0.69}$       &                   & $3.49 \pm 0.55 $     \\[0.1cm]
      29           & \textbf{362},\textbf{987},\textbf{9741}                    & $0.649^{+0.017}_{-0.016}$    & $1.24^{+0.38}_{-0.35}$        & $217 \pm 11$ & $2.41 \pm 0.38$ \\[0.1cm]
      29           & \textbf{362},\textbf{987},\textbf{9741},43142,77424        & $0.676^{+0.016}_{-0.014}$    & $0.74^{+0.22}_{-0.20} $       &                   & $2.51 \pm 0.39 $            \\[0.1cm]
      52           & \textbf{124},\textbf{627},\textbf{8269}                    & $1.490^{+0.036}_{-0.038}$     & $0.94^{+0.65}_{-0.56}$       & $313 \pm 16$ & $1.84 \pm 0.30 $ \\[0.1cm]
      52           & \textbf{124},\textbf{627},\textbf{8269},14723,81049        & $1.511^{+0.049}_{-0.049}$     & $0.60^{+0.40}_{-0.37}$        &                   & $1.87 \pm 0.31 $   \\[0.1cm]
      \textbf{445} & \textbf{1764},\textbf{5104}                                & $0.01793^{+0.00092}_{-0.00091}$ & $0.0207^{+0.0019}_{-0.0018}$ & $86 \pm 8$   & $1.06 \pm 0.31 $ \\[0.1cm]
      \textbf{445} & \textbf{1764},\textbf{5104},71031                          & $0.01903^{+0.00085}_{-0.00090}$ & $0.0223^{+0.0020}_{-0.0016}$ &                   & $1.12 \pm 0.33 $ \\[0.1cm]
      \hline
    \end{tabular}
    \tablefoot{Identification numbers of perturbers and their test asteroids used in each test case. Asteroids whose numbers are written in boldface are included in the DR2. The volume-equivalent diameters correspond to weighted averages taken of multiple previous studies as found in the SiMDA database \citep{Kre20}. The displayed bulk densities were calculated based on the results with the DR2 data included, and the uncertainties correspond to the $1\sigma$ interval.}
  \end{center}
\end{table*}

\label{edna_cogshall}

As described above, Edna had a close encounter with (1764) Cogshall on 31 October 2014 \citep{Gal02}, which corresponds to approximately the middle of the DR2 observational time span. Astrometry of both asteroids is thus included in the DR2 from both before and after the close encounter. It is one of the few cases for which it is possible to attempt mass estimation with the DR2 astrometry alone. Relying on a single test asteroid is not ideal, but it nonetheless serves as an interesting test case.

\begin{figure} 
  \begin{center} 
    \includegraphics[width=1.0\columnwidth]{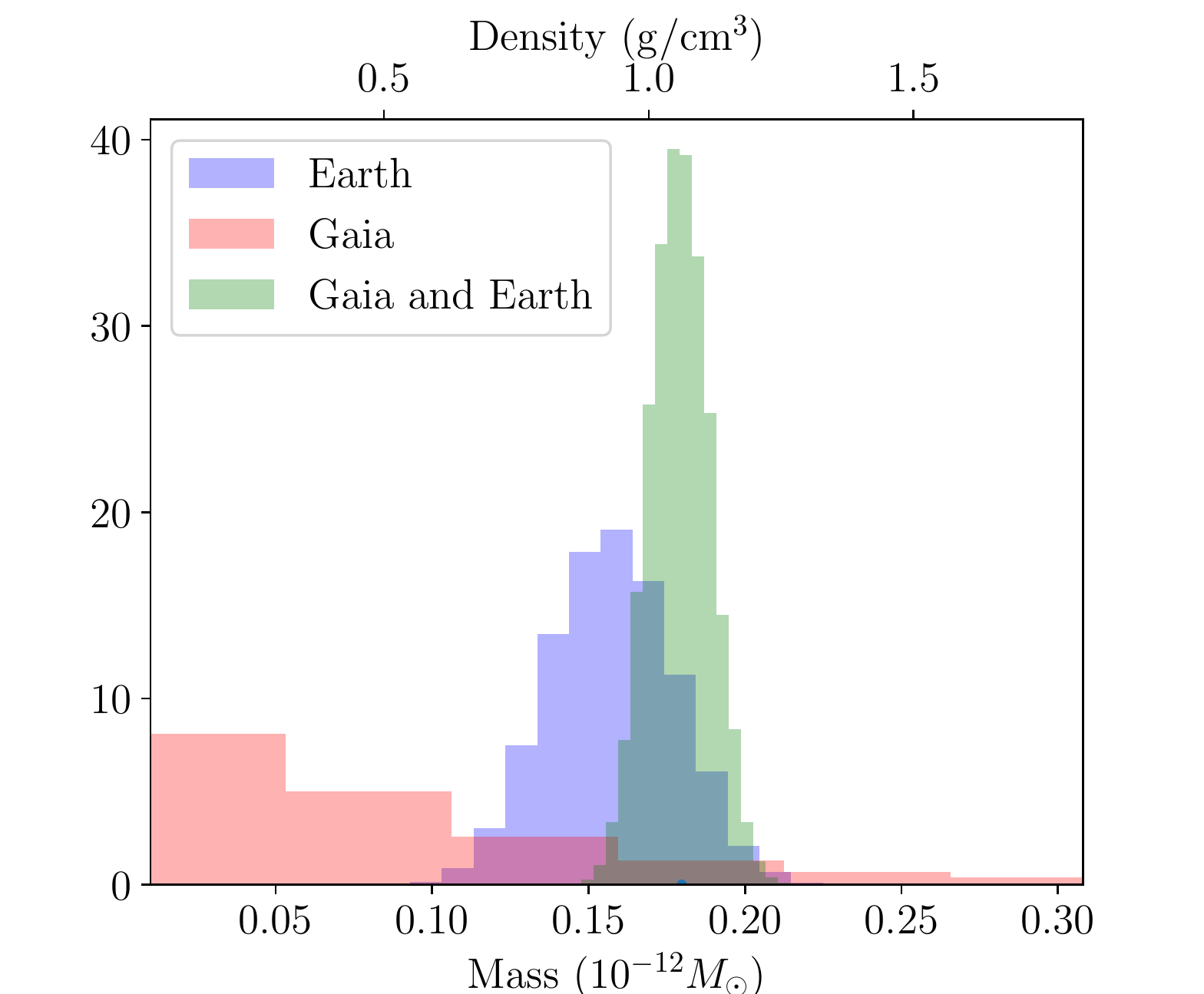} 
    \caption{Probability distributions for the mass of (445) Edna based on its close encounter with (1764) Cogshall with the DR2 astrometry alone (red), the Earth-based astrometry alone (blue), and a combination of the two (green). We note that the densities on the upper $x$ axis do not account for the uncertainty of the volume, which was computed from a volume-equivalent diameter of 86.30 km \citep{Kre20}. Due to the very long tail of the probability distribution in the DR2 case, a large portion of the distribution is beyond the scale of the figure. The frequency on the $y$ axis is normalized such that the integral of each curve is unity.}
    \label{edna}
  \end{center}
\end{figure}

We performed three separate mass estimation runs for (445) Edna with (1764) Cogshall as the sole test asteroid: (i) with the DR2 astrometry only, (ii) with the Earth-based astrometry only, and (iii) with a combination of the DR2 and the Earth-based astrometry. The resulting probability distributions for the mass of Edna are shown in Fig.~\ref{edna}. Upon visual inspection, it is immediately apparent that the mass based on DR2 alone ($2.67^{+10.35}_{-2.67} \times 10^{-14} \msun$) is practically unconstrained and therefore scientifically useless: the maximum-likelihood mass approaches zero, and the uncertainty on the mass is substantial. The runs with the Earth-based astrometry are much better constrained. The Earth-based astrometry alone leads to a maximum-likelihood mass of $1.62^{+0.21}_{-0.21} \times 10^{-13} \msun$, which corresponds to a bulk density of $0.96 \pm 0.30$ g/cm$^3$ computed using a volume-equivalent diameter of $(86.3 \pm 8.28)$ km \citep{Kre20}. The combination of the DR2 and the Earth-based astrometry yields the best result in terms of the uncertainty, and the maximum-likelihood mass of $1.791^{+0.096}_{-0.094} \times 10^{-13} \msun$ results in a bulk density of $1.06 \pm 0.31$ g/cm$^3$.

Mass estimates for Edna found in the literature include $(1.75 \pm 0.39)  \times 10^{-12} \msun$ \citep{Fie10}, $(2.1 \pm 0.6) \times 10^{-12} \msun$ \citep{Gof14}, and $(1.59 \pm 0.79)  \times 10^{-13} \msun$ \citep{Fie20}. These masses correspond to bulk densities of $10.34 \pm 3.76$ g/cm$^3$, $12.41 \pm 5.03$ g/cm$^3$, and $0.93 \pm 0.54$ g/cm$^3$, respectively. The first two values are clearly physically unrealistic because they are higher than the density of pure iron. Our result is in line with that of \citet{Fie20} but has clearly lower uncertainties and thus appears to be the most accurate mass estimate for Edna so far.

\begin{figure} 
  \begin{center} 
    \includegraphics[width=1.0\columnwidth]{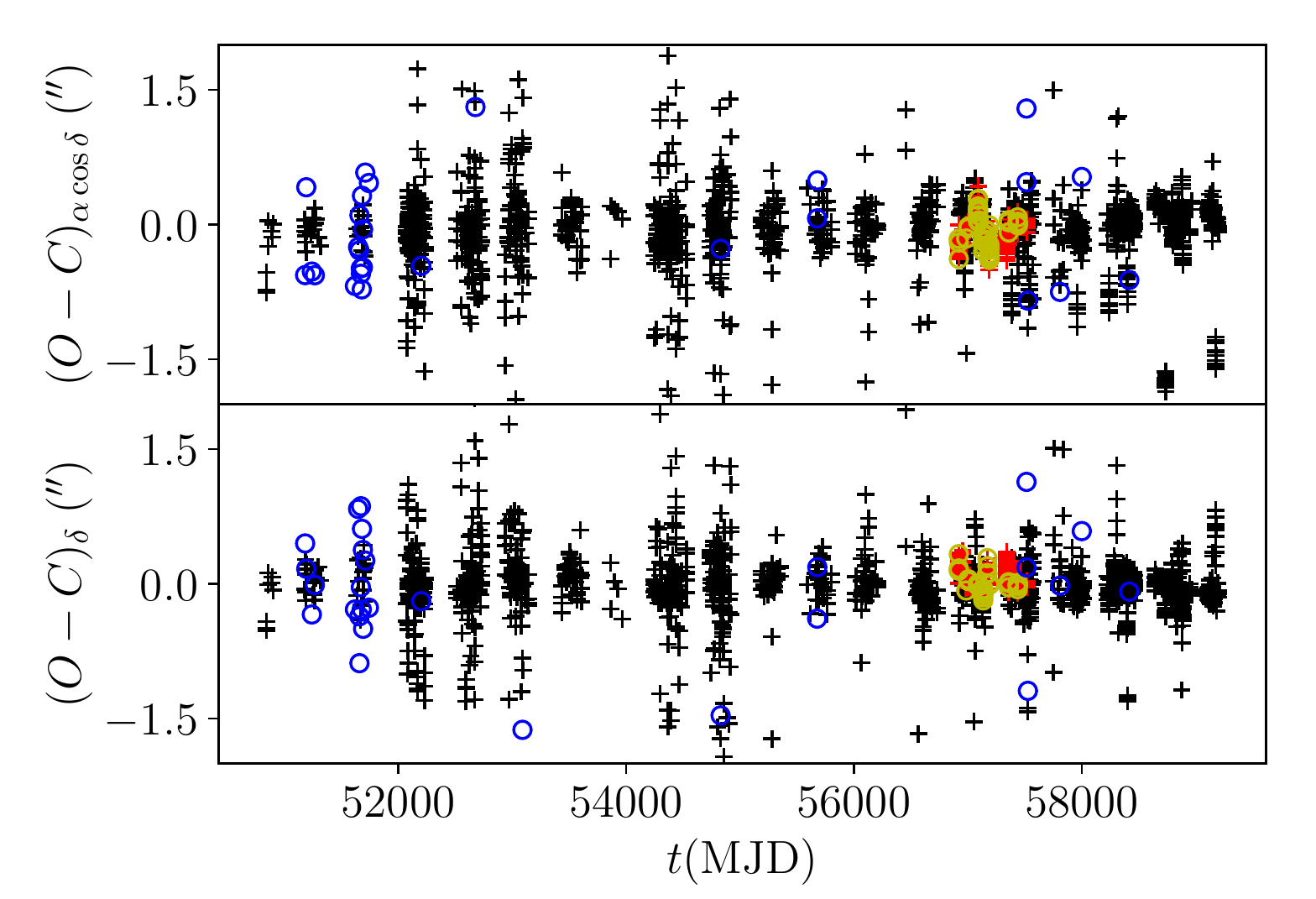}
    \caption{Residuals of the DR2 and the Earth-based astrometry for (445) Edna corresponding to the maximum-likelihood MCMC solution, computed using a combination of the DR2 and the Earth-based astrometry, as a function of modified Julian date. The black data points represent the Earth-based astrometry, the red data points represent the DR2 astrometry, and the blue and yellow circles respectively represent the Earth-based and the DR2 astrometry rejected as outliers. Several data points are not visible due to the scale chosen for the $y$ axis. Consequently, some outlier observations in particular may only be visible in one of the two coordinates.}
    \label{edna_resids}
  \end{center}
\end{figure}

\begin{figure} 
  \begin{center} 
    \includegraphics[width=1.0\columnwidth]{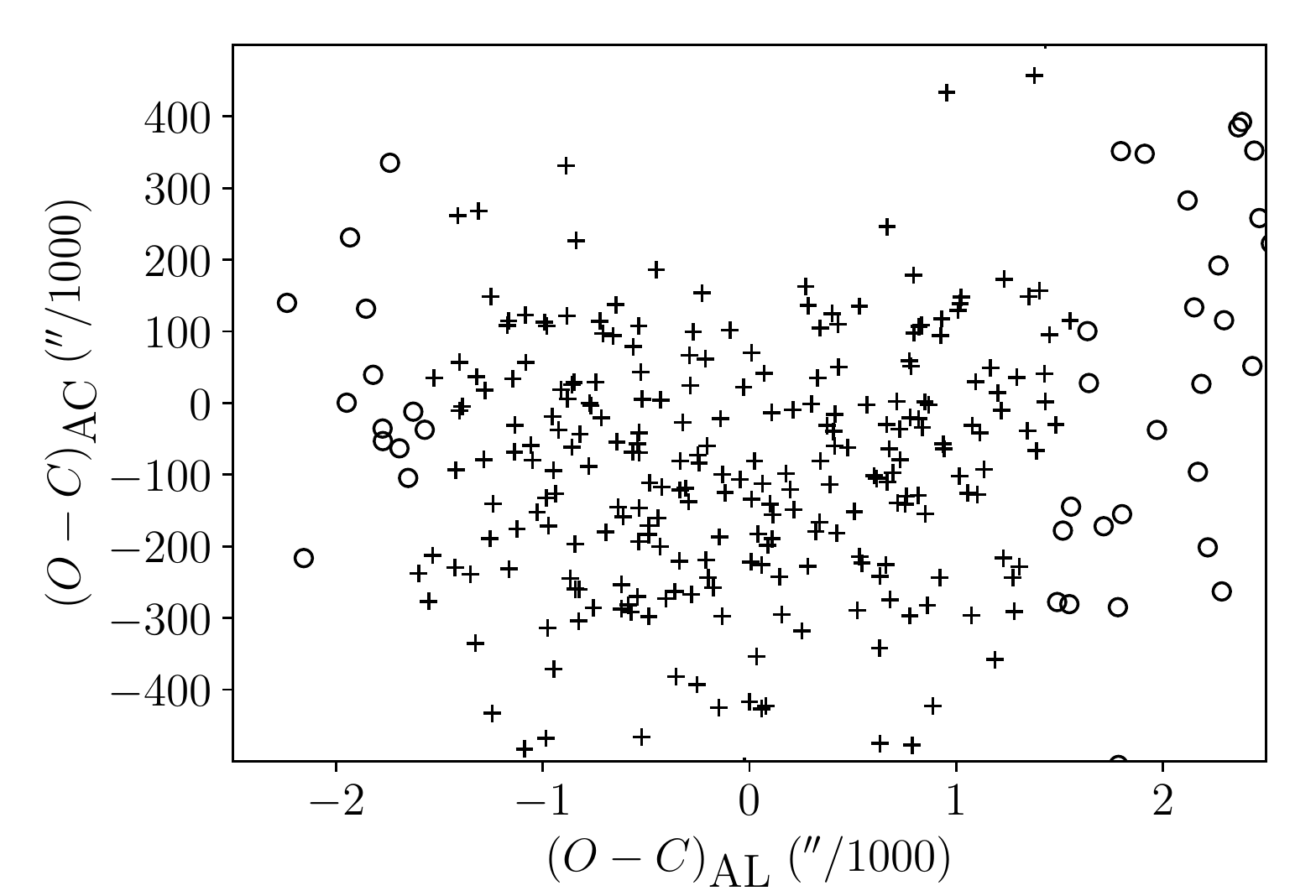}
    \caption{Residuals of the DR2 astrometry for (445) Edna corresponding to the maximum-likelihood MCMC solution, computed using a combination of the DR2 and the Earth-based astrometry, in terms of AL and AC. The unfilled circles represent data rejected as outliers.}
    \label{edna_resids_acal}
  \end{center}
\end{figure}

\begin{figure} 
  \begin{center} 
    \includegraphics[width=1.0\columnwidth]{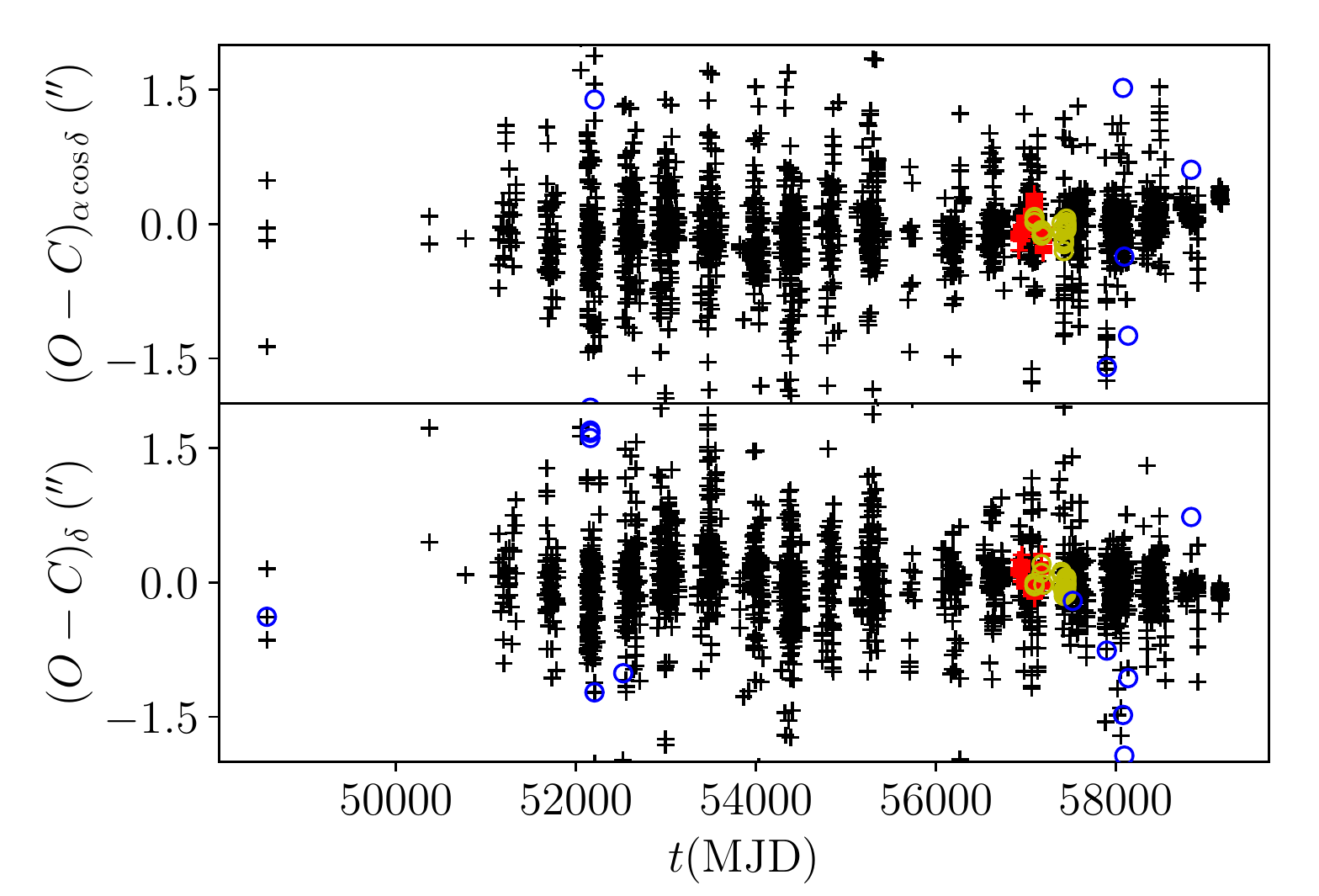}
    \caption{As in Fig.~\ref{edna_resids} but for (1764) Cogshall.}
    \label{cogshall_resids}
  \end{center}
\end{figure}

\begin{figure} 
  \begin{center} 
    \includegraphics[width=1.0\columnwidth]{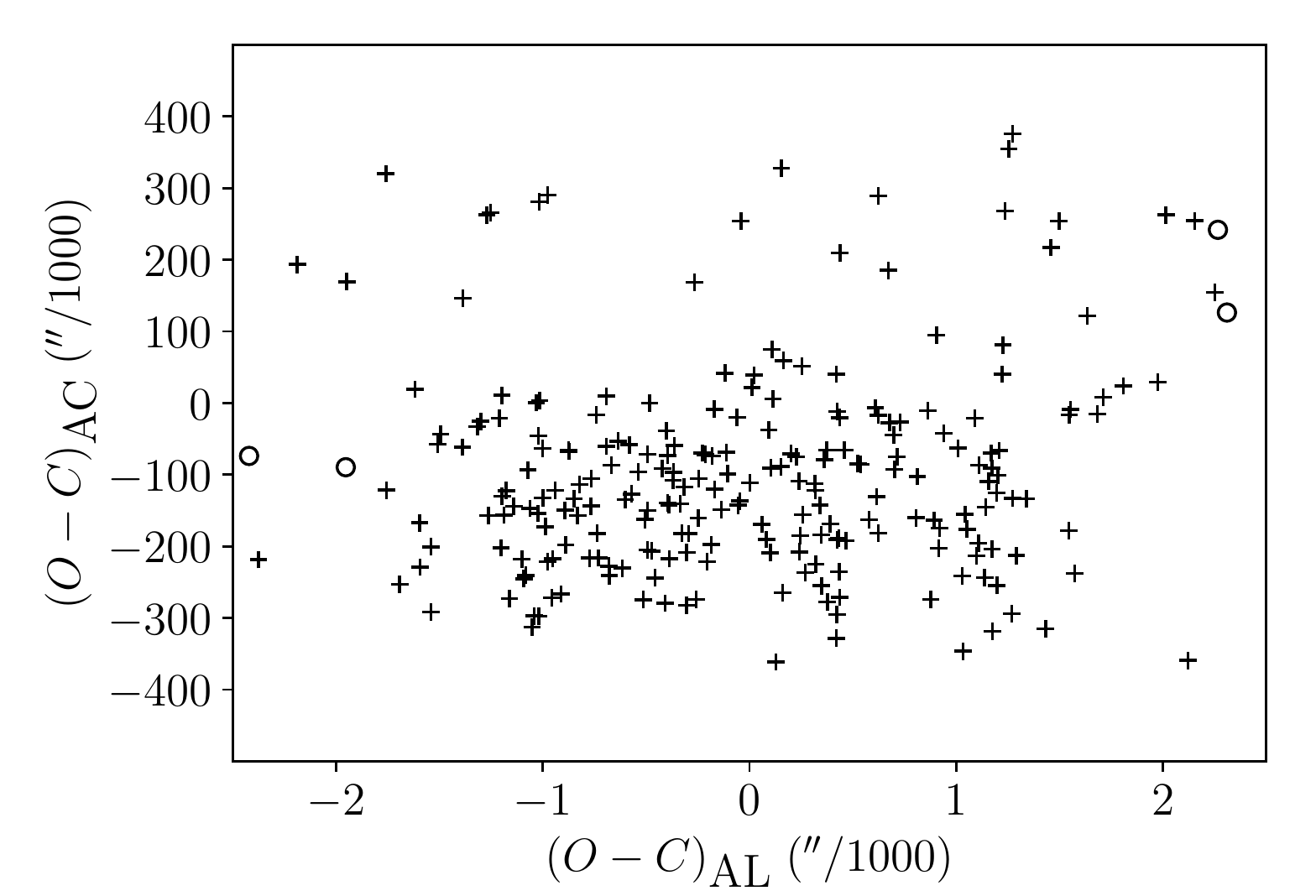}
    \caption{As in Fig.~\ref{edna_resids_acal} but for (1764) Cogshall.}
    \label{cogshall_resids_acal}
  \end{center}
\end{figure}

Figure~\ref{edna_resids} shows the RA and Dec residuals of the maximum-likelihood solution for (445) Edna when combining the DR2 and the Earth-based astrometry. Figure~\ref{edna_resids_acal} shows the DR2 residuals for Edna transformed into the (AL,AC) plane, and Figs.~\ref{cogshall_resids} and \ref{cogshall_resids_acal} show the same for Cogshall. As with (367) Amicitia, in both cases the slight systematic bias toward negative residuals can be seen for the DR2 astrometry in AC but not in AL, which is a known feature of the DR2, as discussed above. Thus, for (445) Edna and (1764) Cogshall, no clear unexpected issues in the residuals can be seen. For the residuals of the DR2 astrometry of Edna, we find a rms value of 200~mas in AC and 0.84 mas in AL, and for Cogshall we find a rms value of 180~mas in AC and 1.0~mas in AL, which are in line with the expected accuracy of the DR2.

With the inclusion of two additional test asteroids, (5104) Skripnichenko and (71031) 1996 EF11, only the first of which is included in the DR2, we find a mass of  $1.903^{+0.085}_{-0.090}\times 10^{-13} \msun$ and a corresponding bulk density of $1.12 \pm 0.33 $ g/cm$^3$ (Table~\ref{results}), demonstrating further improvement on the previous result based on (1764) Cogshall as the sole test asteroid. The difference in the uncertainty is minor, but the maximum-likelihood mass has shifted to a slightly higher value. 

Edna is classified as a Ch-type asteroid, and its density is in line with other such asteroids of similar size \citep[Fig.~9]{Car12}. For C-complex asteroids, there exists a correlation between bulk density and diameter: larger asteroids tend to possess a greater bulk density, which is thought to be caused by the smaller asteroids having porous interiors due to the internal pressure being insufficient for silicate compaction.\ This thus explains Edna's higher porosity leading to a lower density \citep{Car12}.
Ch-type asteroids are considered to be most closely linked to CM chondrites \citep{Riv15}, and the latter have a mean density of 2.2 g/cm$^3$ \citep{Mac11}. Under the assumption that the composition of Edna is similar to CM chondrites, we applied Eq.~2 from \citet{Car12} to compute the corresponding macroporosity:

\begin{equation}
    \mathcal{P} = 100\% \left(1 - {\rho \over \rho_m}\right)
,\end{equation}
where $\rho$ represents the derived bulk density of the asteroid and $\rho_m$ the bulk density of the corresponding meteorites.
Thus, for Edna we find a macroporosity of ($49\pm15$)\%. This result appears entirely realistic.



\subsection{Mass and density of (15) Eunomia}
\label{eunomia}

In the case of (15)~Eunomia, the result with the DR2 and three test asteroids can be viewed with a certain degree of skepticism because the trace of the chain revealed poor mixing. Poor mixing implies a difficulty in generating acceptable proposals, but the root cause is still unclear. The inclusion of two additional test asteroids corrected the issue, and it did not occur in the other reported test cases. The results in terms of mass and density are shown in Table~\ref{results}. For five test asteroids and a combination of the DR2 and the Earth-based astrometry, we find a mass of $(1.522^{+0.047}_{-0.051}) \times 10^{-11} \msun$ for Eunomia (Table~\ref{results}), whereas the average literature value of the mass for this asteroid is $(1.52 \pm 0.16) \times 10^{-11} \msun$ \citep{Kre20}. The result is thus well within expectations. In comparison to the results with the Earth-based astrometry alone, we see an order-of-magnitude reduction in the uncertainty.

Spectroscopically, (15)~Eunomia is an S-type asteroid and is, in fact, the most massive asteroid of its class. As with the C complex, for the S complex there exists a correlation between bulk density and diameter: larger asteroids tend to possess a greater bulk density, which is thought to be caused by the smaller asteroids having porous interiors due to the internal pressure being insufficient for silicate compaction \citep{Car12}. When compared to the relation \citep[Fig.~9]{Car12}, Eunomia's derived bulk density of $3.49 \pm 0.55 $ g/cm$^3$ is in line with expectations of asteroids of its diameter.

S-type asteroids are traditionally associated with ordinary chondrites \citep{Dun10}, the densities of which are very similar to Eunomia's bulk density \citep{Car12}. This suggests that Eunomia has little macroporosity and, consequently, a largely intact structure.

\subsection{Mass and density of (29) Amphitrite}
\label{amphitrite}

For (29)~Amphitrite, the inclusion of two additional test asteroids significantly improved the results in comparison to the case with three test asteroids when the Earth-based astrometry alone was considered. The improvement was much more modest in the DR2 case (Table~\ref{results}). This is easily explained by the three test asteroids in the DR2 having a significantly higher influence on the overall fit in comparison to the asteroids with the Earth-based astrometry alone. Our result of  $0.676^{+0.016}_{-0.014} \times 10^{-11} \msun$ is again in line with the mean literature value of$(0.67 \pm 0.13) \times 10^{-11} \msun$ \citep{Kre20}. In comparison to the results with the Earth-based data alone, an order-of-magnitude reduction in the uncertainty is seen as with (15)~Eunomia.

As was the case with (15)~Eunomia, Amphitrite is an S-type asteroid. When compared to the density-diameter relation of \citet[Fig.~9]{Car12}, Amphitrite's density of $2.51 \pm 0.39$ g/cm$^3$ is in fact anomalously low, with the expected density being roughly 40\% higher. A previous analysis of Amphitrite's spectra suggests that it is best represented by primitive achondrites such as lodranites or winonaites \citep{Hir90}.

Lodranites have an average bulk density of $3.53$ g/cm$^3$ \citep{Mac11}, which corresponds to a macroporosity of $(29 \pm 11)\%$ for Amphitrite when combined with our bulk density estimate. The average bulk density for winonaites is listed as $3.24$ \citep{Mac11}, which corresponds to a macroporosity of $(23 \pm 12)\%$.

\begin{figure}  
  \begin{center} 
    \includegraphics[width=1.0\columnwidth]{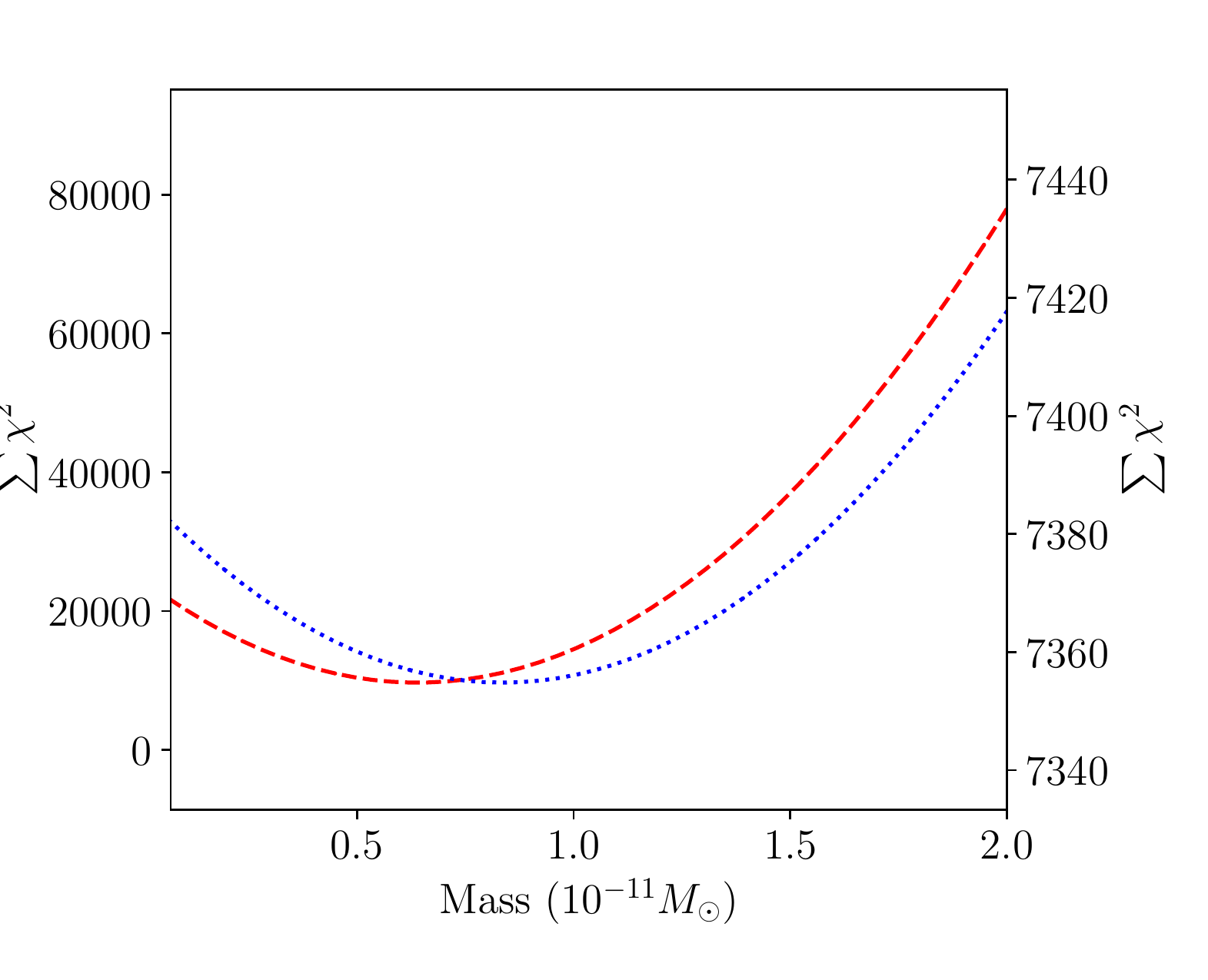}
    \caption{Total $\chi^2$ of all objects as a function of mass obtained from the marching mass estimation algorithm applied to (29) Amphitrite with (dashed red line) and without the DR2 astrometry (dotted blue line). The left $y$ axis represents the run with the DR2 and the Earth-based astrometry, and the right axis represents the run with the Earth-based data alone.}
    \label{29_march}
  \end{center}
\end{figure}

As an additional test to verify the significant reduction of the uncertainty when including the DR2 astrometry and to gauge the impact of Amphitrite's mass on the goodness of fit in terms of $\chi^2$, we applied the marching approximation to (29) Amphitrite, with all five test asteroids and their corresponding maximum-likelihood orbits determined by the MCMC algorithm. The results are depicted in Fig.~\ref{29_march}, which shows that the impact on the mass on the overall goodness of fit is indeed enormous in the DR2 case in comparison to the case without. As the acceptance probability does not allow for moves toward significantly worse $\chi^2$ values, one can easily imagine that the range of possible masses allowed by the MCMC chain should indeed be extremely small in the DR2 case. In fact, the allowed range is within the step size (in mass) used in the marching algorithm. So, the width of the underlying probability distribution is expected to be smaller than the difference between individual tested masses, which is found to be the case. In comparison, with the Earth-based astrometry alone, the range in acceptable mass is significantly larger (Table~\ref{results}). Thus, in general it appears that when \gaia{} astrometry is used for some test asteroids, there is little reason to include test asteroids without \gaia{} astrometry in the computations due to their influence on the goodness of fit being comparatively small. We conclude that this test provided results in line with our expectations.

\subsection{Mass and density of (52) Europa}
\label{europa}

In the case of (52)~Europa, the inclusion of two additional test asteroids in fact slightly worsened the results (Table~\ref{results}), which may be explained by deficiencies in our force model for the aforementioned asteroids, such as perturbations caused by other close encounters that were unaccounted for. This time, the result of $(1.511^{+0.049}_{-0.049}) \times 10^{-11} \msun$ is also above the mean value of $(1.28 \pm 0.41) \times 10^{-11} \msun$ \citep{Kre20} but well within its uncertainties. Recent studies, such as that of \citet{Bae17}, also find results strongly in line with ours. Once again, we observe an order-of-magnitude reduction in the uncertainty compared to the results with the Earth-based data alone.

Europa is classified as a C-type asteroid, and thus, as was the case with (445)~Edna, it is part of the C-complex. We find a density of $1.87 \pm 0.31$ g/cm$^3$ for Europa, which is in line with expected values based on its diameter \citep[Fig.~9]{Car12}. There are currently no known meteorite matches for Europa in particular \citep{Tak14}, but, for example, \citet{Car12} links C-type asteroids in general with CM chondrites that have a mean bulk density of 2.2 g/cm$^3$. That assumption yields a macroporosity of $(17 \pm 14)\%$ for Europa.

\section{Conclusions}

Based on our study, it is clear that the inclusion of \gaia{} astrometry can lead to order-of-magnitude improvements in the uncertainties of the masses of asteroids. We observe that, while traditionally the uncertainty of the mass has been the dominant term in the corresponding uncertainty of the bulk density \citep{Kre20}, with the accuracy enabled by the DR2 data the uncertainty of the volume is starting to dominate the uncertainty budget for bulk density. This applies, in particular, to large asteroids with significant amounts of astrometric data.

However, the DR2 is still partly handicapped by both the relatively small number of asteroids included and the short time span covered. We found that mass estimation with the DR2 alone appears to be currently unfeasible. However, in the case of (445)~Edna, a combination of the DR2 and the Earth-based astrometry leads to significantly lower uncertainties than the Earth-based astrometry alone. These results further lead to a realistic bulk density with lower uncertainties than any prior estimates. Thus, the DR2 astrometry must be combined with Earth-based astrometry to result in useful mass estimates for asteroids. The forthcoming Gaia DR3 is expected to include a significantly larger number of asteroids, in addition to more astrometry for those that were already included in the DR2. It will thus allow for mass estimation for a much larger sample of asteroids and further improve the estimated masses of asteroids already studied here. We conclude that the future of asteroid mass estimation looks very bright indeed.

\begin{acknowledgements}
This work was supported by grants \#299543, \#307157, and \#328654 from the Academy of Finland. This research has made use of NASA's Astrophysics Data System, and data and/or services provided by the International Astronomical Union's Minor Planet Center. The computations for this work were mainly performed with the Kale cluster of the Finnish Grid and Cloud infrastructure (persistent identifier: urn:nbn:fi:research-infras-2016072533). 
We thank Paolo Tanga and Federica Spoto for their assistance with the astrometric debiasing and weighting schemes, respectively, and Grigori Fedorets for the OpenOrb implementation of his normal points algorithm for Gaia astrometry, which proved useful for debugging purposes. We thank Francois Mignard for graciously sharing his Fortran code on the Gaia relativistic light bending correction with us. Finally, we thank the anonymous referees for their comments, which helped to improve the paper.
\end{acknowledgements}

\bibliographystyle{aa}
\bibliography{refs}

\end{document}